%% file: main.tex
\documentclass[conference]{IEEEtran}
\usepackage{graphicx}
\usepackage{amsmath}
\usepackage{amssymb}
\usepackage{mathtools}
\usepackage{bm}
\usepackage{subfig}
\usepackage{wrapfig}
\usepackage{comment}
\usepackage{xcolor}

\usepackage{array}
\newcolumntype{P}[1]{>{\centering\arraybackslash}p{#1}}
\newcolumntype{M}[1]{>{\centering\arraybackslash}m{#1}}

\newcommand{\ie}{{\em i.e.}}
\newcommand{\eg}{{\em e.g.}}

\def\rrr#1\\{\par
\medskip\hbox{\vbox{\parindent=2em\hsize=6.12in
\hangindent=4em\hangafter=1#1}}}

\begin{document}

\input{frontmatter-sv}

\begin{abstract}
\input{abstract}

\end{abstract}

\section{Introduction}
\label{sec:introduction}
\input{introduction}

\section{Imbalance and Computation Complexity in Traffic Engineering}
\label{sec:problem}
\input{problem}

\section{Taming the imbalance to Improve TE Solution}
\label{sec:solution}
\input{solution}

\section{Performance Evaluation}
\label{sec:evaluation}
\input{evaluation}

\section{Related Work}
\label{sec:related}
\input{related}

\section{Conclusions}
\label{sec:future}
\input{future}

\section*{Acknowledgments}
This material is based upon work supported by the National Science Foundation under Grant No. OAC-2029278.

\bibliographystyle{IEEEtran}
\bibliography{./reconfiguration, ./resilience}


\end{document}

%% file: frontmatter-sv.tex
\title{Taming Imbalance and Complexity in WAN Traffic Engineering}
\thanks{This work was supported by the
    US National Science Foundation under Grant OAC-.
}
\author{\IEEEauthorblockN{Yufeng Xin, Sajith Sasidharam, Cong Wang, and Mert Cevik}
\IEEEauthorblockA{
RENCI, University of North Carolina at Chapel Hill\\
Chapel Hill, NC, USA\\
}
}

\maketitle
\thispagestyle{empty}

%% file: abstract.tex
The rapid expansion of global cloud infrastructures, coupled with the growing volume and complexity of network traffic, has fueled active research into scalable and resilient Traffic Engineering (TE) solutions for Wide Area Networks (WANs). Despite recent advancements, achieving an optimal balance between solution quality and computational complexity remains a significant challenge, especially for larger WAN topologies under dynamic traffic demands and stringent resource constraints.

This paper presents empirical evidence of a critical shortcoming in existing TE solutions—their oversight in adequately accounting for traffic demand heterogeneities and link utilization imbalances. We identify key factors contributing to these issues, including traffic distribution, solver selection, resiliency, and resource overprovisioning. To address these gaps, we propose a holistic solution featuring new performance metrics and a novel resilient TE algorithm. The metrics—critical link set and network criticality—provide a more comprehensive assessment of resilient TE solutions, while the tunnel-based TE algorithm dynamically adapts to changing traffic demands.

Through extensive simulations on diverse WAN topologies, we demonstrate that this holistic solution significantly improves network performance, achieving a superior balance across key objectives. This work represents a significant advancement in the development of resilient and scalable TE solutions for WANs.

%% file: introduction.tex
In recent years, the adoption of centralized traffic engineering (TE) control by major Cloud and Internet Service Providers (ISPs) has been pivotal in enhancing network throughput while mitigating network congestion~\cite{singh2022traffic}. This approach's promise lies in its capacity to optimize flow routing and distribution, dynamically accommodating varying traffic demands and network states within short timeframes. TE controllers operate cyclically within predetermined reconfiguration time windows, typically five to fifteen minutes. In each window, a series of complex tasks are initiated from the TE controller: predicting traffic demands, employing a TE optimizer to find solutions, and updating network flow routings in the devices  accordingly~\cite{bogle2019teavar,TE:Contracting:NSDI21}.

Despite ongoing advancements, two major challenges continue to complicate the optimization of TE control, often introducing conflicting requirements. The first challenge lies in balancing computational complexity with the optimality of TE models and solutions. For realistically sized WANs, solving traditional linear or mixed-integer programming models within the short re-optimization windows required for effective TE control becomes impractical with current optimization solvers. The second challenge pertains to balancing the increasing traffic demands with maintaining network resilience in the face of potential failures. Existing solutions often resort to excessive resource overprovisioning, leaving much of the network capacity idle, which leads to inefficient utilization.
 
Recent TE solutions have favored a ``tunnel-based" approach, allowing demand splitting across a limited set of precomputed paths between node pairs. This approach can significantly reduce the complexity of the resulting Mathematical Programming model, compared to the alternative ``link-based" approach. However, the run time still grows super-linearly with the size of the network even for the simplest feasible flow solution~\cite{TE:Contracting:NSDI21}. The tunnel configuration streamlines routing updates by redistributing demands solely among tunnels, thus circumventing the need for updating flow rules across all routers along the paths. Importantly, this approach naturally addresses concerns regarding network resilience, as demands can be redistributed among available tunnels upon failure(s)~\cite{kang2015efficient}. 

Enhanced network availability is attainable through diverse resilience schemes employing proactive (``protection") or reactive (``restoration") recovery procedures when network failures occur. Traditionally categorized into ``path-based" and ``link-based," these schemes vary in their coverage of network failures (e.g. single or multiple link and/or node failures, or shared risk link groups - SRLG).

Notably, a state-of-the-art approach pursues ``congestion-free" solutions~\cite{wang2010r3, liu2014ffc, jiang2020pcf}. This optimization strategy aims to distribute traffic across the network without causing link oversubscription under all targeted failure conditions. However, the worst-case nature of ``congestion-free" protection, treating all possible failures equally, results in heightened computation complexity and underutilization of backup resources, which leads to significantly less amount of satisfied traffic demands. Recent studies propose a relaxation of these requirements by limiting the failure set's scope by integrating the probabilities of network element failures into the formulation~\cite{bogle2019teavar,chang2019lancet,zhong2021arrow}.

Conversely, another line of solutions focuses on restoration by rapidly recomputing TE after failure detection, redistributing demands among available tunnels~\cite{TE:Contracting:NSDI21, xu2023teal}. Although flows may sustain losses during recomputation, this approach yields improved TE performance without significant under-provisioning of demands.

Nevertheless, these solutions have only studied the average link utilization without exploring the global network capacity utilization distribution. Our empirical analysis reveals several more insightful observations. Firstly, the majority of the links are seriously under-utilized 
in common TE optimization solutions. Secondly, a large number of pre-configured tunnels may not be used depending on the optimization solver of choice. 
Last but not least, failures of links with very low utilization would not affect satisfying the overall traffic demand.  In summary, WAN demonstrates 
strong demand imbalance and high link and tunnel utilization heterogeneity in the typical TE solutions.  

These observations lead us to develop a holistic TE solution featuring new performance metrics and a novel resilient TE algorithm. The metrics—critical link set and network criticality—provide a more comprehensive assessment of network imbalances, while the tunnel-based TE algorithm dynamically adapts to traffic demand heterogeneities. Our approach effectively reduces resource wastefulness and computation complexity while retaining superior resilient TE performance.

This paper proceeds as follows: Section~\ref{sec:problem} defines the base resilient TE optimization problem, illustrating demand imbalances, link utilization heterogeneity, and other factors impacting the overall solution quality. In Section~\ref{sec:solution}, we introduce a novel network criticality metric and propose a TE solution employing varying numbers of preconfigured tunnels. Section~\ref{sec:evaluation} presents performance evaluations through extensive numerical analyses on realistic network topologies and traffic demands. Section~\ref{sec:related} offers a concise summary of existing work, while Section~\ref{sec:future} concludes the paper with future work.

%% file: problem.tex
In this section, we present the fundamental formulation of the tunnel-based resilient Traffic Engineering (TE) solution. We demonstrate the problem of highly imbalanced traffic demands and link utilization, and excessive resource overprovisioning in exchange for resilience using a typical WAN topology. Additionally, we show that existing solutions lead to low tunnel utilization and high computational time, highlighting the need for more efficient resilient TE solutions.

\subsection{Tunnel-Based TE Optimization}
Adhering to the definition and formulation presented in~\cite{liu2014ffc} that have been used by most recent TE solutions. 
A wide-area network is modeled as a bi-directional graph $G(V,E)$  with nodes $V$ connected by links $E$. 
Each link $e(u, v) \in E$, with its start node $u$ and end node $v$, possesses properties like maximum available bandwidth capacity  $c_e$ and a cost function $w_e$ 
to be used in the optimization objective function. 

The traffic demands are defined as a set of commodities $F$, where each $f$ represents a demand $f(s,t)$ between source node 
$s \in V$ and destination node $t \in V$, with bandwidth requirement $d_f$. These demands are structured in a traffic matrix (TM). 
A demand $f(s,t)$ is routed through a predefined set of tunnels (paths) 
$T_f$ from the node $s$ to the node $t$ in the network represented by a binary variable 
$x_{e, t}$, indicating if edge $e$ is part of path $t \in T_f$.    

The TE problem is formulated as a constrained maximum network flow problem with link capacity constraints $(2)$, flow distribution constraints $(3)$, and demand constraints $(4)$. It can be seen that this model will always admit an optimal solution on the flow variables $b_f$, but the solution may not satisfy all the demands ($d_f$).

\setcounter{equation}{0}
\begin{align}
\max &\sum_{f \in F} b_f
\end{align}
\text{\it s.t.} 
\setcounter{equation}{1}
\begin{align}
	& \sum_{f \in F, t \in T_f}a_{f, t} \cdot x_{e, t}  \leq c_e, \quad \forall e \in E \\
	& \sum_{t \in T_f} a_{f, t}  \geq b_f, \quad \forall f \in F \\
	& 0 \leq b_f \leq d_f; 0 \leq a_{f, t},  \quad \forall f \in F,  \forall t \in T_f 
\end{align}

After each TE run, a feasible solution decides the variable $a_{f, t}$, the portion of demand $f$ assigned to tunnel $t$, referred to as a flow. Demands distributed over multiple flows can be implemented using routing rules installed in the tunnel routers~\cite{split:kang2015}.

\subsection{Resilient TE}
We consider a set of $Q$ resilience scenarios. For the more popular single link failure case, $Q=E \bigcup \{0\}$. A $q \in Q$, could represent a link failure or the normal condition when $q=0$. We define an identity variable $y_e^q$ whose value is $1$ when link $e$ fails under failure scenario $q$, $0$ otherwise. $y_e^q = 1$ indicates that a feasible tunnel can not be created using the link $e$. This implies that the step to pre-compute the tunnels needs to enumerate all possible $y_e^q$ values for each demand.  

To achieve ``congestion-free'' under any failure scenario, 
the tunnel set $T_f$ in the constraints in the base TE model needs to be substituted with $T_{f}^{q}$, $\forall q \in Q$ that represents the available tunnels in $T_f$ under failure scenario $q$. 

This strategy was first proposed as the ``Forward Fault Correction'' (FFC) scheme~\cite{liu2014ffc}. However, covering all failures in the TE formulation significantly escalates computational complexity because computational complexity is jointly decided by the size of the topology, the number of tunnels, and the number of failures. 

All recent protection solutions fell in between the two solutions, via pruning the covered failures based on the link availability measures in order to reduce the computation time and bandwidth overprovisioning~\cite{chang2019lancet,bogle2019teavar}. The most recent solution~\cite{xu2023teal} leverages the fast inference capability of pre-trained machine learning models to minimize the TE re-computation time, essentially a fast restoration scheme. Its main caveat is that the trained model may not cover demand profiles well. More importantly, none of these TE resilience solutions take the demand heterogeneity into consideration.

In the rest of this section, we present a detailed analysis of the TE solution using the Google B4 WAN topology and an associated benchmark demand matrix that has been used as a baseline in all the recent studies~\cite{xu2023teal}. By default, the number of pre-computed tunnels per node pair is set to five in all scenarios and only one link failure is considered in the FFC model. Based on the analysis, we derive several key findings and identify key factors affecting the solution quality and computational time.

\subsection{Demand Heterogeneities}

\begin{figure}[!ht]
\begin{center}
\vspace{-0.1in}
\includegraphics[width=0.52\textwidth]{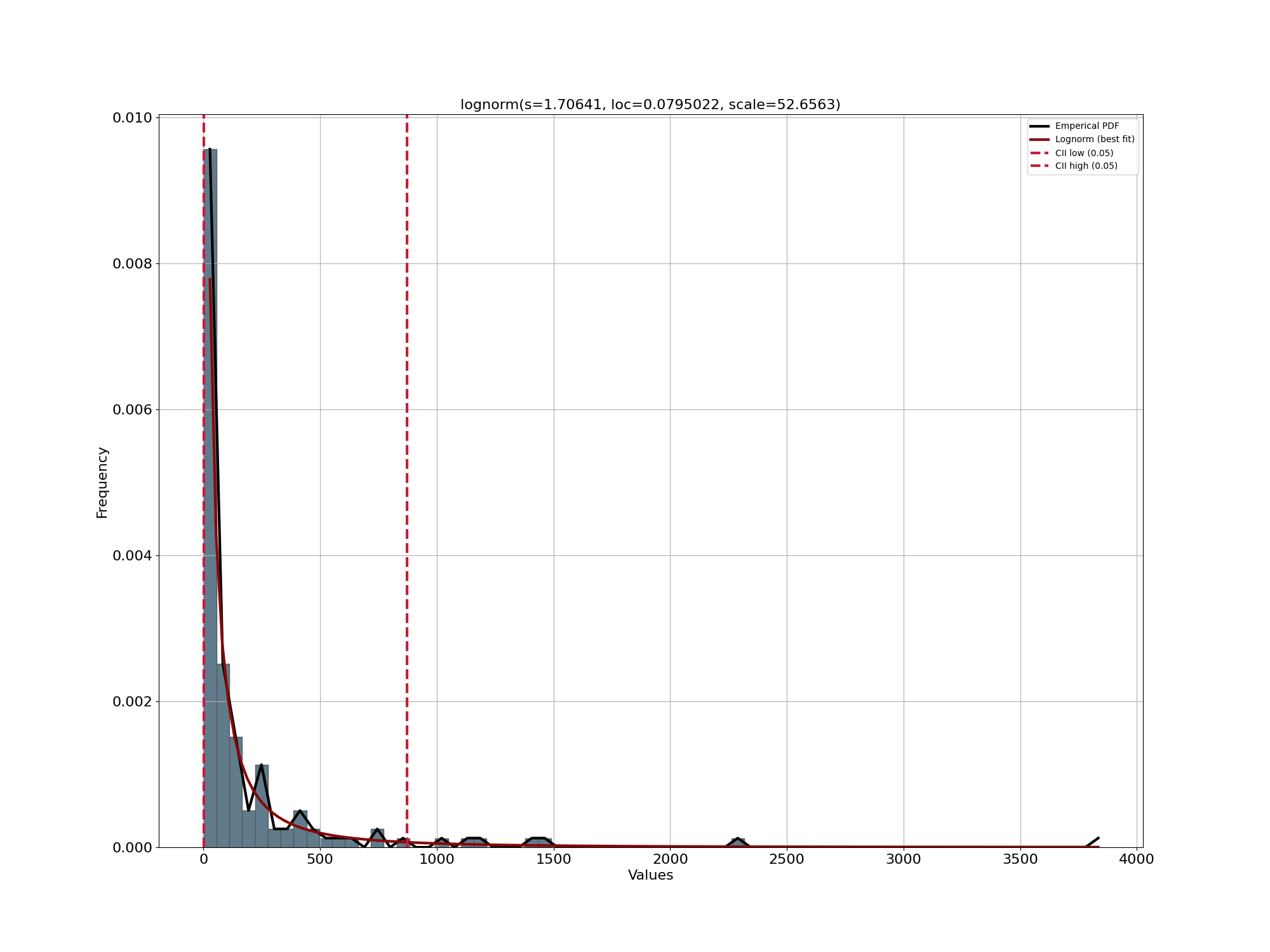}
\end{center}
\vspace{-0.05in}
\caption{Lognormal Distribution Fitting on B4 Demand}
\label{fig:b4:demand}
\end{figure}

Internet traffic volume typically follows a long-tail lognormal distribution, with a small number of flows demanding majority bandwidth in a TM~\cite{traffic:lognormal:ToN21}. 
The benchmark B4 traffic matrix aligns well with a lognormal distribution, whose PDF (Probability density function) fitting is shown in Fig.~\ref{fig:b4:demand}. Notably, the majority 
of traffic lie on the lower end of the bandwidth demand distribution.
 
The inherent long-tail distribution of traffic demands is a key factor driving link utilization imbalances in TE solutions, as will be elaborated in the following sections.

\subsection{Performance Metrics}
We evaluate several performance metrics to gain comprehensive insights into TE solutions.
\begin{itemize}
\item{Solver time:} This metric captures the computational time required for a solver to find an optimal solution for a given traffic matrix. 
It excludes one-time operations such as tunnel path computation and the formulation of the optimization model in the solver.
\item{Mean utility:}  It measures the average capacity utilization across all network links.
\item{Overprovisioning Ratio:} This metric measures the total flows in the FCC solution over the total demands in the given traffic matrix, capturing the ``congestion free'' cost for TE resilience.
\item{Unmet flow ratio:} This ratio accounts for the total flows in the optimal solution, \ie, the result of the objective function (1), to the total requested bandwidth 
in the traffic matrix. It normalizes the total flow against the total demand.
\item{Unmet demands ratio:} This is defined as the percentage of traffic demands whose provisioned total flows fall short of the requested amount in the input traffic matrix.
\item{Used Tunnel Ratio:} This metric quantifies the proportion of pre-computed tunnels that carry assigned flows in the optimal solution, measuring how efficiently tunnel resources are utilized.
\end{itemize}

\subsection{Solver Selection}
We evaluate two popular optimization solvers: Google OR-Tools (using the default GLOP optimizer)~\cite{Google:or} and the CVXPY package (employing the default LP optimizer) commonly used in existing solutions. While both can find the optimal objective values for both TE and FFC models, their solutions and computation times differ dramatically. This is primarily due to the different algorithms they use by default, Simplex or Interior-point method. 

Table~\ref{tab:b4:unmet} compares the performance of two solvers on the B4 topology in terms of demand satisfaction, overprovisioning ratio, and average link utilization.

GLOP uses significantly shorter computation time. Compared to CVXPY, GLOP uses less computation time by at least two orders of magnitude.
GLOP also leaves the majority of tunnels unused and incurs lower link utilization.  On the other hand, CVXPY results in a slightly smaller number of unmet 
demands and distributes flows to tunnels evenly. This is significant in two aspects: computation time is the key metric for TE and high link utilization 
will lead to lower resilience and robustness. In addition, as we will show in the evaluation section, CVXPY often fails to obtain a solution for bigger networks due to its high memory requirements. 

\begin{table}[!h]
\begin{center}
    \begin{tabular} { | p {1.25 cm} | p {1 cm} | p {1 cm} | p {1 cm} |  p {1 cm} | p {1 cm} |}
    \hline
     {\scriptsize Solver} & {\scriptsize Unmet Demands} & {\scriptsize Over-provisioning} & {\scriptsize Mean Utility}  & {\scriptsize Used Tunnels} & {\scriptsize Solver Time}  \\
    \hline
    TE(GLOP) & 0 & 0 & 19.8\% & 20.1\% & 0.006 \\
    FFC(GLOP) & 6.8\%  & 73.8\% & 41.9\%  & 40.6\% & 0.173 \\
    TE(cvxpy) & 0 & 0  & 30.5\% & 100\% & 0.71   \\
    FFC(cvxpy) & 3.79\%  & 189.4\%\ & 74.5\%  & 96.3\% & 9.27  \\
    \hline
    \end{tabular}
\caption{Optimization Solver}
\label{tab:b4:unmet}
\end{center}
\end{table}

\subsection{Link Utilization Imbalance}
A major limitation observed in existing studies is their focus on average or overall performance metrics, such as maximum flow or average network utilization, 
while largely neglecting the distribution of link utilization. Some links can be highly underutilized, while other links are fully utilized.
    
\begin{figure}[!ht]
\begin{center}
\includegraphics[width=0.45\textwidth]{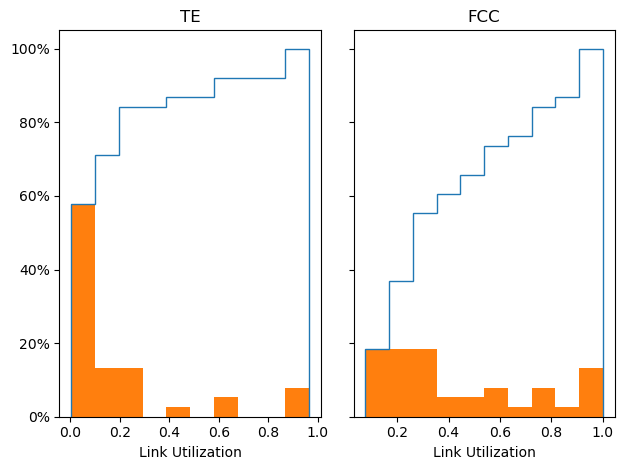}
\end{center}
\vspace{-0.05in}
\caption{Link Utilization Distribution on B4 Topology}
\vspace{-0.05in}
\label{fig:b4:utilization}
\end{figure}

Figure~\ref{fig:b4:utilization} illustrates the imbalance of link utilization distributions in the TE and FFC solutions obtained from the two solvers. 

In the TE solution, most links are significantly underutilized, with nearly $60\%$ of them operating at less than $30\%$ capacity, and only a small portion of links being heavily utilized. 
The FFC solution, on the other hand, flattens the link utilization distribution to some degree, as it reserves additional spare capacity to accommodate potential failures.

\subsection{Overprovisioning and Demand Satisfaction}
Table~\ref{tab:b4:tunnel} highlights the high resource overprovisioning underscoring the congestion-free FCC solutions. 
We compare the performance of TE and FFC models across two solvers, GLOP and CVXPY, under different tunnel configurations (T=3 and T=5).

The FFC model incurs significant overprovisioning compared to the TE model which can satisfy all the demands without overprovisioning, 
especially when using the CVXPY solver. For instance, with T=3, the FFC model's overprovisioning reaches $197.6\%$. 
As the tunnel count increases to T=5, the overprovisioning for FFC slightly decreases in CVXPY ($189.4\%$) 
but increases in GLOP (from $64.9\%$ to $79.7\%$). At the same time, the FFC model also results in higher link utilization 
compared to the TE model.

A notable improvement is observed in the reduction of unmet demand as the number of tunnels increases, particularly in the FFC model. 
For GLOP, unmet demand reduces from $15.9\%$ at T=3 to $6.8\%$ at T=5, while CVXPY shows a decrease from $7.57\%$ to $3.79\%$.

\begin{table}[!h]
\begin{center}
    \begin{tabular} { | p {1.25 cm} | p {0.75 cm} | p {0.75 cm} | p {0.75 cm} |  p {0.75 cm} | p {0.75 cm} |  p {0.75 cm} | }
    \hline
    & \multicolumn{2} { | c | } {Unmet Demand} & \multicolumn{2} { | c | }{Overprovisioning}  & \multicolumn{2} { | c | }{Mean Utility}\\
    \hline
    & T=3 & T=5 & T=3  & T=5 & T=3  & T=5 \\
    \hline
    TE(GLOP) & 0 & 0 & 0 & 0 & 19.8\% & 19.8\% \\
    FFC(GLOP) & 15.9\%  & 6.8\% & 64.9\%  & 79.7\% & 35.1\% &44.3\% \\
    TE(cvxpy) & 0 & 0 & 0 & 0 & 25.7 & 30.5\% \\
    FFC(cvxpy) & 7.57\%  & 3.79\%\ & 197.6\%  & 189.4\% & 67\% &74.5\% \\
    \hline
    \end{tabular}
\caption{Number of Pre-configured Tunnels}
\label{tab:b4:tunnel}
\end{center}
\end{table}

In summary, ``congestion-free" resilient TE solutions like FFC sacrifice demand satisfaction and result in high network utilization in exchange for resiliency 
via excessive network resource overprovisioning. Factors like demand distribution, number of pre-configured tunnels, and optimization 
solver selection contribute to the resilient TE solution quality and feasibility.

%% file: solution.tex
To address the resource over-provisioning and computation complexity challenges in existing TE solutions, we developed novel enhancements to take advantage of the inherited imbalance in traffic demand and network utilization.

\subsection{Adaptive Pre-Computed Tunnels}
The study presented in the last section shows that the number of pre-computed tunnels is crucial to both solution quality and computational complexity. The observed imbalance in flow distribution among tunnels suggests that decreasing the number of tunnels in the model might not sacrifice the solution quality while reducing computation time. However, such a reduction may compromise resilient TE performance as available tunnels may become insufficient in post-failure scenarios. Previous studies universally employed homogeneous tunnels, typically four or five~\cite{xu2023teal}. To achieve a better tradeoff, we propose an adaptive tunnel scheme in which more tunnels would be allocated to larger demands while fewer tunnels are sufficient for smaller demands.

Leveraging the long-tail distribution pattern observed in WAN traffic, predominantly consisting of small demands, we design a TE algorithm incorporating varying tunnels based on demanded volume. This design involves defining a function $T_f$ in the TE model (Section~\ref{sec:problem}.A) that determines the value of $T_f$ based on the demand $d_f$:

\begin{equation}
T_f = T(d_f) 
\end{equation}

This adaptive scheme will assign a smaller number of pre-configured tunnels to most demands and consequently significantly reduce variables and computation time in the resulting TE optimization model. To ensure the existence of spare tunnels under failures, $T_f$ should always be greater than one. 

We note that, though the tunnel computation is a one-time operation, configuring and managing many tunnels is an enormous burden to the WAN operators. Our proposed scheme will significantly reduce the operational cost. 

\subsection{Network Criticality}
The uneven distribution of traffic flows across network links suggests that certain links might be more critical than others in facilitating overall traffic demands. The links that carry most flows can substantially impede network flows when congested or broken. To quantify link importance under a given traffic matrix, inspired by the percolation theory in network science~\cite{hamedmoghadam2021percolation}, we define a criticality score for each link in a TE solution as follows:

\begin{equation}
S_{e} =\sum_{f \in F, e=e^{*} \in T_f} \frac{|b_f|}{|F|}
\end{equation} 

where $e^{*}$ is the most loaded link on $P_f$, \ie, $U_{e^{*}} \ge U_e, \forall e \in T_f$, and $U_e$ is the utilization of link $e$. 
The links with higher criticality scores can be regarded as the {\it bottlenecks} for a specific traffic matrix.    

We then define a unified {\it network criticality} metric $R$ that associates link utilization and criticality, addressing demand and link utilization altogether:

\begin{equation}
R =\sum_{e \in E} \frac{S_{e}}{U_{e}}
\end{equation} 

This metric reflects the capacity of a network and the capability of a TE solution to satisfy traffic demands while withstanding link failures and congestion
 under traffic fluctuation with more rooms in the overall link capacity. More satisfied demands or lower link utilities will lead to higher network criticality.
 Lower $R$ indicates the network is subject to higher congestion situations.

Table~\ref{tab:b4:criticality} shows the percentage of critical links among all links and the network criticality for both scenarios.

\begin{table}[!h]
\begin{center}
{\scriptsize
    \begin{tabular} { | p {1.1 cm} | p {0.35 cm} | p {0.45 cm} | p {0.45 cm} | p {0.45 cm} | p {0.4 cm} |  p {0.4 cm} | p {0.4 cm} |  p {0.4 cm} | }
    \hline
    Solver & \multicolumn{2} { | c | } {Unmet Flow} & \multicolumn{2} { | c | } { Used Tunnels} & \multicolumn{2} { | c | }{ \#Critical links}  & \multicolumn{2} { | c | }{ R}\\

    & T=3 & T=5 & T=3 & T=5 & T=3  & T=5 & T=3  & T=5 \\
    \hline
  TE(GLOP) & 0 & 0 & 33.3\% & 20..2\% & 42.1\% & 29\% & 6.626 & 3.565 \\
   FFC(GLOP) & 15\% & 14.1\% & 59.9\%  & 40.6\% & 44.7\%  & 36.8\% & 6.197 & 4.407 \\
   TE(cvxpy) & 0 & 0 & 100\% & 100\% & 52.6\% & 26.3\% & 7.08 & 1.75 \\
    FFC(cvxpy) & 15\% &  14.1\% & 100\%  & 96.4\% & 39.5\%  & 26.3\% & 7.37 &3.612 \\
    \hline
    \end{tabular}
\caption{Critical Links and Network Criticality}
\label{tab:b4:criticality}
}
\end{center}
\end{table}

%% file: evaluation.tex
In this section, we present the evaluation results focusing on the scalability of the TE solutions with respect to the scales of the network and traffic matrix. All experiments were conducted on the Longleaf cluster, a Linux-based computing environment at the University of North Carolina at Chapel Hill. The cluster servers boasted Intel processors clocked at $2.3 - 2.5$ GHz, cache sizes of $24.75 - 30$M, and RAM capacities ranging from $256$ GB to $754$ GB. 

We evaluate the TE solutions on two representative WAN topologies, a smaller one and a bigger one, under varied traffic loads with both GLOP optimizer and CVXPY solver. The smaller Google's private WAN (B4) topology comprises $12$ nodes and $38$ edges, and the bigger UsCarrier topology spans $158$ nodes and $378$ edges~\cite{xu2023teal}. UsCarrier demonstrates greater sparsity and larger diameter compared to the B4 topology.

We precomputed up to five shortest paths between each node pair as candidate tunnels for flow allocation in the TE model.  
For the adaptive tunnel scheme, based on extensive simulation studies with different combinations, all demands in a traffic matrix are sorted ascendingly on their bandwidth requirement and then evenly divided into three groups that are assigned three, four, or five pre-computed tunnels accordingly. 

A traffic demand trace from B4~\cite{singh2022traffic} is used as the baseline traffic matrix for the B4 topology. The traffic matrix for the UsCarrier topology is generated from the lognormal distribution we fitted from the B4 demand matrix (Section~\ref{sec:problem}.C). Link capacities were increased for the UsCarrier topology to satisfy this base demand matrix. 

For each topology, we vary the demand matrix scales by $0.5$ to $2$ times and evaluate all the six performance metrics plus the network criticality for the TE and FFC models with fixed (curve legend names ending with a $\_0$) and adaptive tunnels (curve legend names ending with a $\_1$). Due to significant disparities among solutions, log scales were used for computational time. 

\subsection{TE Optimization}

\begin{figure}[!ht]
    \subfloat[Computation Time\label{b4:te:time}]{%
      \includegraphics[width=0.23\textwidth]{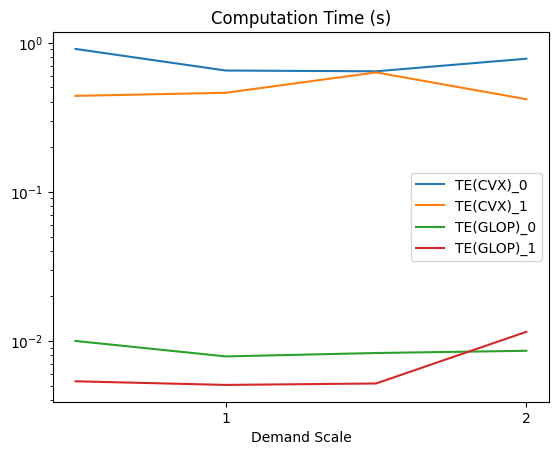}
    }
    \hfill
    \subfloat[Mean Utility\label{b4:te:mean}]{%
      \includegraphics[width=0.23\textwidth]{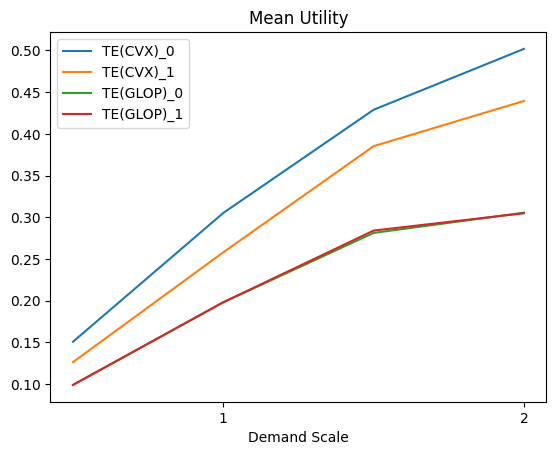}
    }
    \hfill
    \subfloat[Used Tunnel Ratio\label{b4:te:over}]{%
      \includegraphics[width=0.23\textwidth]{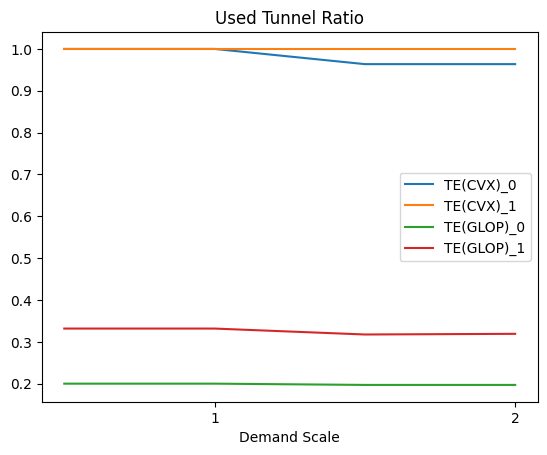}
    }
    \hfill
    \subfloat[Unmet Flow Ratio\label{b4:te:unmetflow}]{%
      \includegraphics[width=0.23\textwidth]{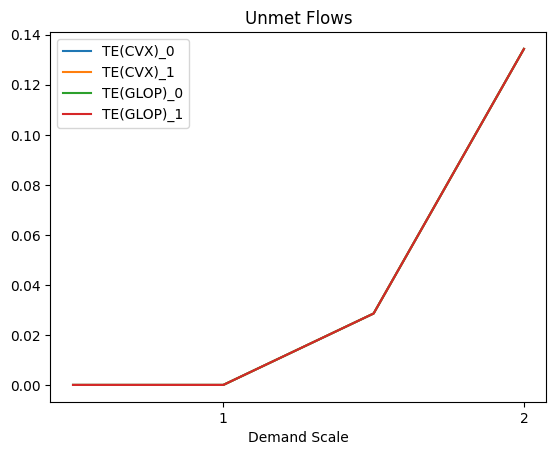}
    }
    \hfill
    \subfloat[Unmet Demands Ratio \label{b4:te:unmetdemand}]{%
      \includegraphics[width=0.23\textwidth]{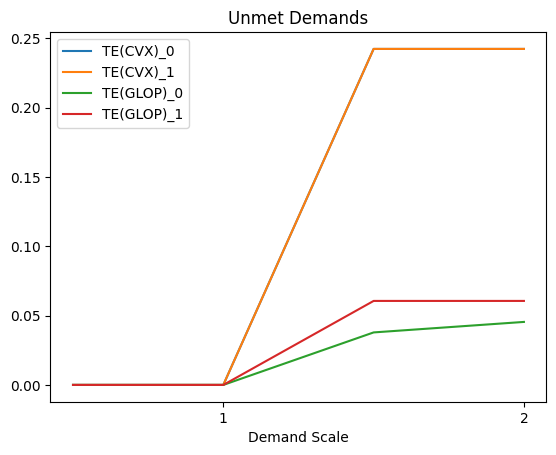}
    }
    \hfill
    \subfloat[Network Criticality\label{b4:te:nc}]{%
      \includegraphics[width=0.23\textwidth]{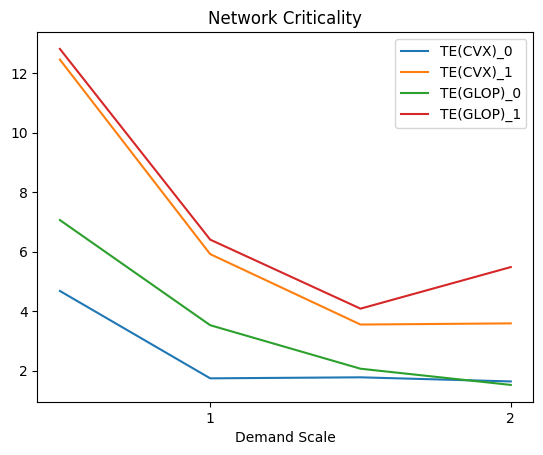}
    }
    \caption{TE Performance: B4 Topology (N=13, E=38)}
    \label{fig:b4:te}
 \end{figure}

The set of figures in Figure~\ref{fig:b4:te} provides a comprehensive view of TE performance on the B4 topology. We scaled up the demand matrix so that some demands can not be satisfied. 
Both GLOP and CVXPY solvers can solve the basic TE problem within sub-seconds, with GLOP running notably faster and resulting in lower link utilization. 

\subsubsection{Small Topology}

With the GLOP solver, the adaptive tunnel scheme further reduces the computational time until the demands are scaled up by two. The mean utility stays almost the same. When the demands scale up ($scale > 1$), unsatisfied demands start showing up. However, the adaptive tunnel scheme does not increase the unmet flow ratio while increasing the unmet demand ratio slightly. 
More positively, it effectively increases tunnel utilization indicated by the used tunnel ratio. 

As a result of the lowered link utilization and unchanged flow satisfaction performance, the proposed adaptive tunnel scheme effectively reduces the number of critical links and therefore higher network criticality performance. We note that $R$ decreases when the traffic demands initially scale up along with the increased unsatisfied demands. However, the adaptive tunnel scheme shows good scalability as the network criticality gradually flattens first and then starts to trend up modestly.   

\subsubsection{Big Topology}
  
\begin{figure}[!ht]
    \subfloat[Computation Time\label{us:time}]{%
      \includegraphics[width=0.23\textwidth]{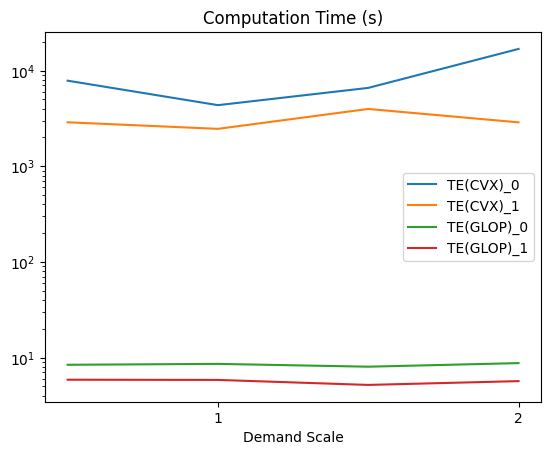}
    }
    \hfill
    \subfloat[Mean Utility\label{us:mean}]{%
      \includegraphics[width=0.23\textwidth]{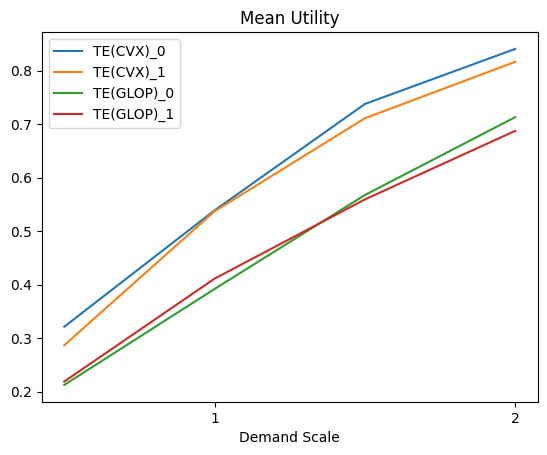}
    }
    \hfill
    \subfloat[Used Tunnel Ratio \label{us:over}]{%
      \includegraphics[width=0.23\textwidth]{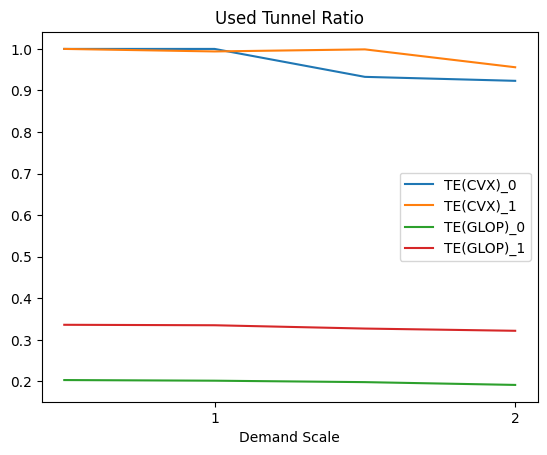}
    }
    \hfill
    \subfloat[Unmet Flow Ratio\label{us:unmetflow}]{%
      \includegraphics[width=0.23\textwidth]{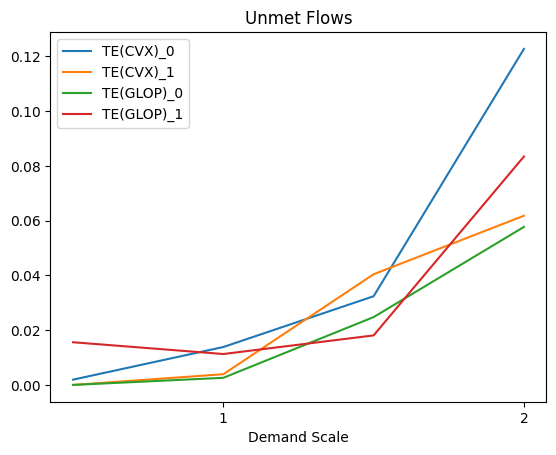}
    }
    \hfill
    \subfloat[Unmet Demands Ratio \label{us:unmetdemand}]{%
      \includegraphics[width=0.23\textwidth]{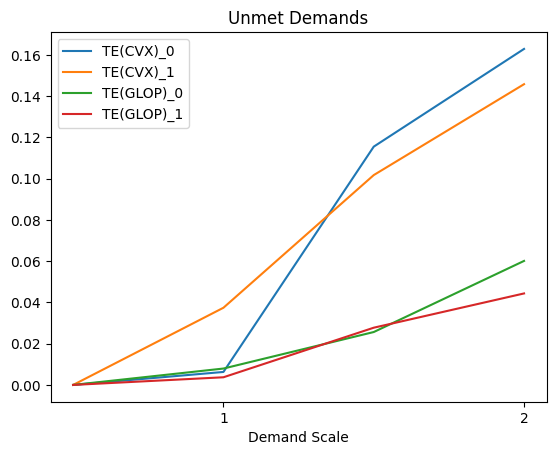}
    }
    \hfill
    \subfloat[Network Criticality\label{us:nc}]{%
      \includegraphics[width=0.23\textwidth]{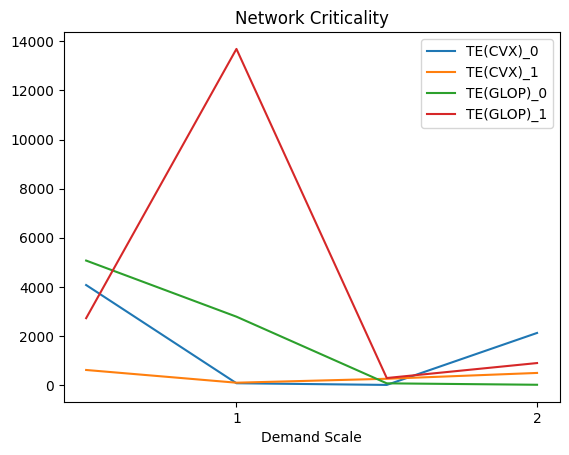}
    }
    \caption{TE Performance: UsCarrier Topology (N=158, E=378)}
    \label{fig:us:te}
 \end{figure}

Figure~\ref{fig:us:te} illustrates the TE performance on the much bigger UsCarrier topology with varying demand scales. The CVXPY solver ($TE(CVX)\_0$) with homogeneous tunnels took hours to compute. With the adaptive tunnel scheme ($TE(CVX)\_1$) the time is significantly reduced though still too high to be useful in reality. When using the GLOP solver, applying the adaptive tunnel scheme reduces the computation time to a few seconds. GLOP also results in lower mean utility, unmet flow ratio, and unmet demand ratio on higher demand scales. Using the adaptive tunnel scheme only affects these three metrics marginally. The used tunnel ratio is favorably increased with a much smaller total number of tunnels to be pre-configured. Applying the adaptive tunnel scheme performs better overall in terms of the metrics, especially at the higher demand scales. Network criticality $R$ reaches a much higher value than the small B4 network as expected due to the sheer size of the network and the traffic demands. It starts to decrease with increasing demand scales due to the heightened link utility and unmet demands (Figure~\ref{fig:us:te} (b) (d) (e)), reflecting increased network congestion. However, it decreases slowly and then starts to trend up, the same as we observed on the B4 network. It also results in higher tunnel utilization (Figure~\ref{fig:us:te} (c)).
  
 \subsection{Resilient TE with FFC}
 FFC aims to achieve {\it congestion free} by covering all the possible failure scenarios. 
 As such, it will incur a much longer computation time compared to the basic TE solution. 
 As reported in~\cite{xu2023teal}, the existing solutions in the FFC family often fail to find solutions for topologies bigger than B4.
 
We already showed empirical evidence that FFC gains resilience through excessive resource overprovisioning. In this section, we present more 
detailed performance implications of FFC and how our proposed schemes can achieve better overall performance.   
 
\subsubsection{Small Topology}

  \begin{figure}[!ht]
    \subfloat[Computation Time\label{b4:ffc:time}]{%
      \includegraphics[width=0.23\textwidth]{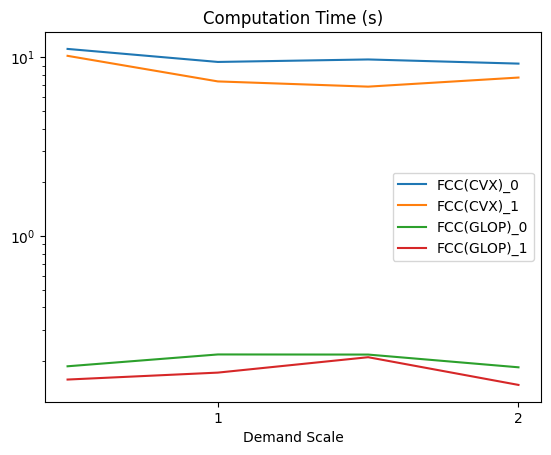}
    }
    \hfill
    \subfloat[Mean Utility\label{b4:ffc:mean}]{%
      \includegraphics[width=0.23\textwidth]{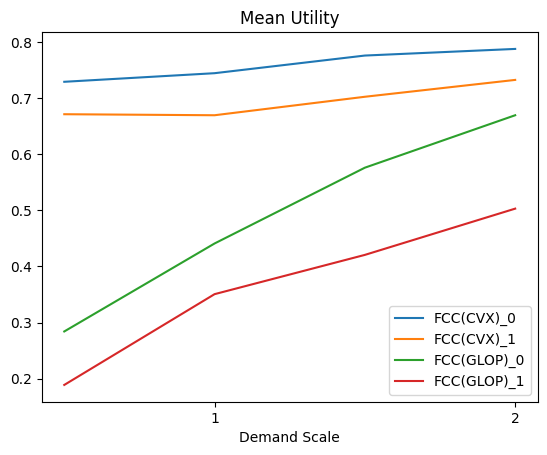}
    }
    \hfill
    \subfloat[Overprovisioning Ratio\label{b4:ffc:over}]{%
      \includegraphics[width=0.23\textwidth]{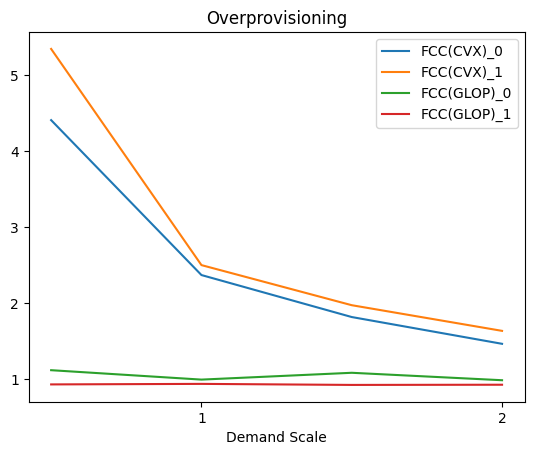}
    }
    \hfill
    \subfloat[Unmet Flow Ratio\label{b4:ffc:unmetflow}]{%
      \includegraphics[width=0.23\textwidth]{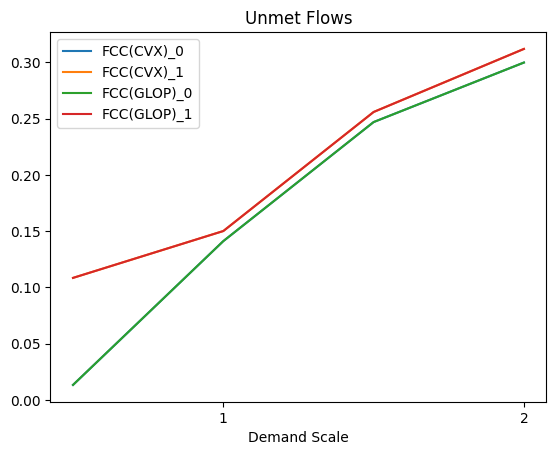}
    }
    \hfill
    \subfloat[Unmet Demands Ratio \label{b4:ffc:unmetdemand}]{%
      \includegraphics[width=0.23\textwidth]{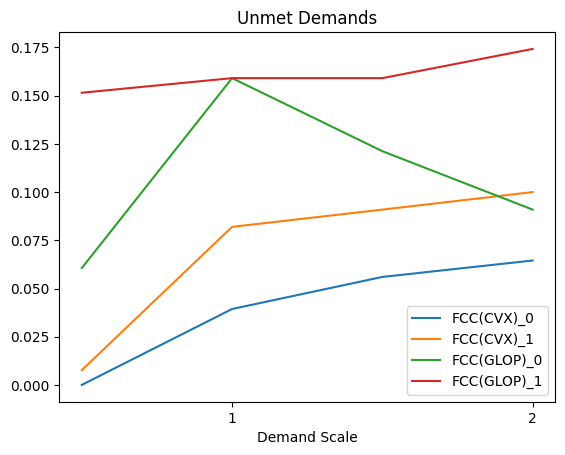}
    }
    \hfill
    \subfloat[Network Criticality\label{b4:ffc:nc}]{%
      \includegraphics[width=0.23\textwidth]{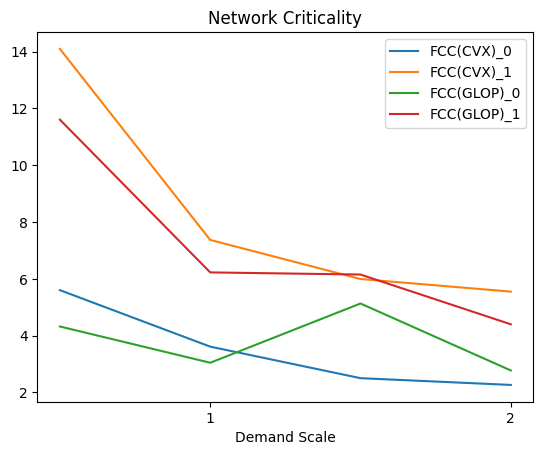}
    }
    \caption{FFC Performance: B4 Topology (N=13, E=38)}
    \label{fig:b4:ffc}
 \end{figure}
 
As shown in Figure~\ref{fig:b4:ffc}, CVXPY finds the optimal solution for the small B4 topology within ten seconds, while GLOP finishes within sub-seconds. GLOP solution achieves 
notably lower mean utility and overprovisioning ratio at the cost of higher unmet demands. Although both found optimal solutions, Overall the GLOP solution leads to higher network criticality. 
In both cases, the adaptive tunnel scheme improves the computation time, mean utility, and overprovisioning performance. It does cause marginal performance deterioration in the category of unmet demands, especially at higher demand scales. 
 
Figure~\ref{b4:ffc:over} clearly shows that FFC relies on extremely high overprovisioning to achieve {\it congestion free} resilience, as much as more than $400\%$ with CVXPY solver. 
The overprovisioning ratio trends lower over increased demand scales only because of the increased unmet traffic demands. 
Coupled with the adaptive tunnel scheme, the GLOP solution effectively reduces the overprovisioning ratio under $100\%$.
 
\subsubsection{Big Topology}
  \begin{figure}[!ht]
    \subfloat[Computation Time\label{us:ffc:time}]{%
      \includegraphics[width=0.23\textwidth]{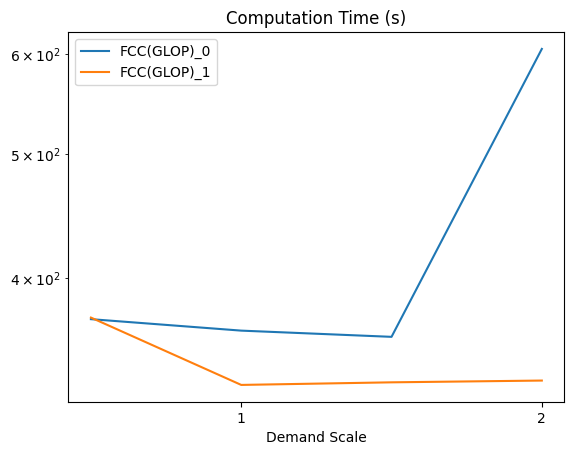}
    }
    \hfill
    \subfloat[Mean Utility\label{us:ffc:mean}]{%
      \includegraphics[width=0.23\textwidth]{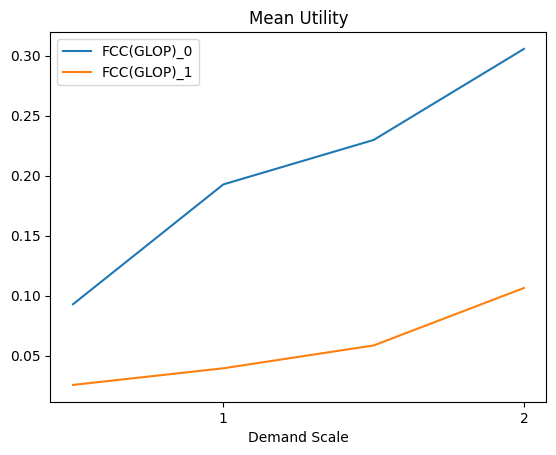}
    }
    \hfill
    \subfloat[Overprovisioning Ratio\label{us:ffc:over}]{%
      \includegraphics[width=0.23\textwidth]{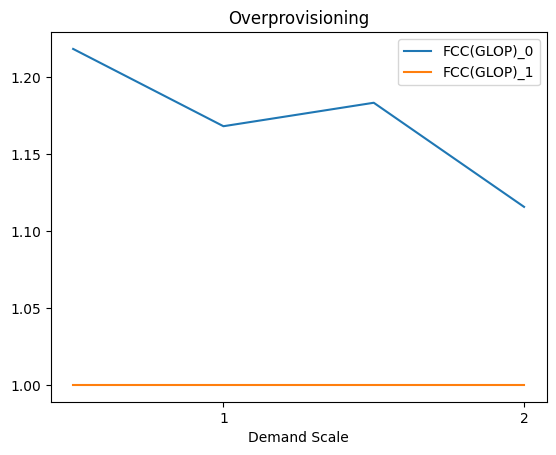}
    }
    \hfill
    \subfloat[Unmet Flow Ratio\label{us:ffc:unmetflow}]{%
      \includegraphics[width=0.23\textwidth]{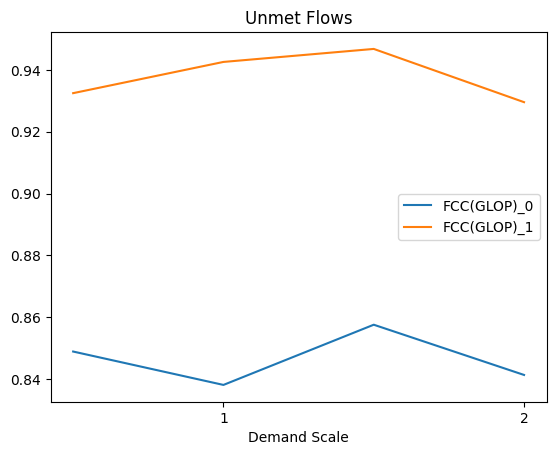}
    }
    \hfill
    \subfloat[Unmet Demands Ratio \label{us:ffc:unmetdemand}]{%
      \includegraphics[width=0.23\textwidth]{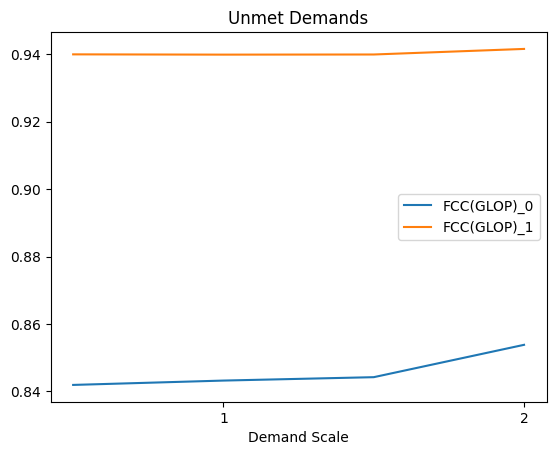}
    }
    \hfill
    \subfloat[Network Criticality\label{us:ffc:nc}]{%
      \includegraphics[width=0.23\textwidth]{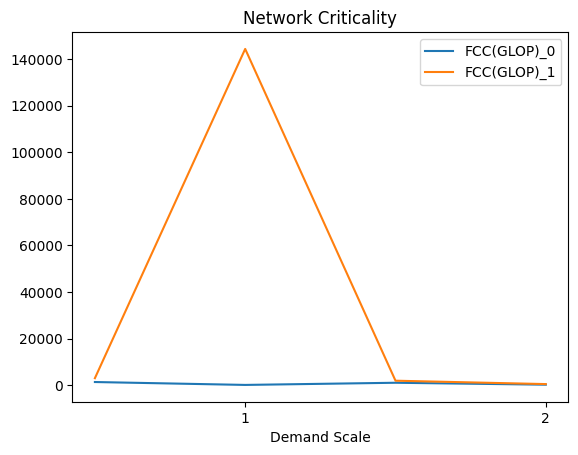}
    }
    \caption{FFC Performance: UsCarrier Topology (N=158, E=378)}
    \label{fig:us:ffc}
 \end{figure}

Figure~\ref{fig:us:ffc} depicts FFC performance on the big UsCarrier topology. Because the CVXPY solver takes a long computational time and eventually fails due to an out-of-memory error, only results from the GLOP solutions are presented. The first observation is that the adaptive tunnel scheme can significantly reduce the computation time, \eg, approximately five minutes when the demand scale is two. It also effectively reduces the overprovisioning ratio. As expected, the FFC solution leads to a high percentage of unmet demands with the high demand volumes in the big network. In comparison, we recall the TE solution can satisfy most demands in the same traffic matrix. The adaptive tunnel scheme moderately lowers the flow and demand satisfaction with a lower link utility. Overall it results in a higher network criticality. The network criticality does show some irregularity as it becomes very high with the original demand scale ($=1$).  

Our main takeaway is that the proposed adaptive tunnel scheme, coupled with the right solver, makes the  {\it congestion free} resilient TE solution feasible for big networks in production settings. 
Besides the extra benefits of managing a much smaller number of pre-configured tunnels, it also leads to higher network criticality performance without significantly compromising other performance metrics, such as the loss of flows and demands.

%% file: related.tex
Traditional Traffic Engineering (TE) relies on extending routing protocols in IP and MPLS networks, utilizing algorithms such as the Constrained Shortest Path (CSP) algorithm or heuristics. However, these methods often result in suboptimal network utilization and slow routing convergence. The MPLS-TE solution also faces the challenge of high cost and operational complexity~\cite{TE:TOTEM07,TE:SDN:CN14}. 

Layer-2 WAN services, such as MPLS, introduce integer constraints for unsplittable demands, leading to Mixed Integer Programming (MIP) formulations, which are computationally intractable. In contrast, modern TE controllers focus on layer-3 services, where traffic flows can be split and under-provisioned, allowing for more flexible traffic provisioning within capacity constraints. 

Software-defined networking (SDN) enables global optimization for an entire traffic matrix through centralized control. Large private cloud SD-WAN solutions now use TE solvers to re-optimize networks periodically in intervals ranging from three to fifteen minutes to accommodate dynamic network states~\cite{b4:after:sigcomm18,swan:sigcomm13}. The multi-commodity flow problem forms the core of these TE models. Tunnel-based models are particularly favored because they result in easier LP (Linear Programming) formulations~\cite{TE:SDN:Infocom13,TE:path:12}. However, the computational time still grows super-linearly with the size of the network, even for simplified maximum flow formulations. As such, reducing computational complexity has become a focal point in recent TE studies. Various TE variants exist, differing broadly in performance implications and computational complexity with different objective functions~\cite{TE:Objective:07}. Proposed techniques to reduce computational complexity include topology abstraction or segmentation~\cite{TE:Contracting:NSDI21,TE:Blastshield:NSDI22}. 

Resilient TE solution has attracted extensive research interest in recent years~\cite{wang2010r3, liu2014ffc, jiang2020pcf}. Unfortunately integrating the resilience requirement into TE optimization further heightens computational complexity and reduces overall traffic flow. The pure ``congestion-free" model will make finding the TE optimization solution infeasible even for networks of moderate size. To address this challenge, recent studies propose relaxation models, limiting the failure set's scope through deterministic or probabilistic techniques to make resilient TE feasible for bigger networks~\cite{bogle2019teavar,chang2019lancet,zhong2021arrow}.

Recently, deep learning methods have been explored to address these challenges, offering scalable and robust solutions for large WANs. A notable example is the {\it Teal} algorithm, which incorporates graph neural networks (GNN), multi-agent reinforcement learning (RL), and distributed optimization to provide near-optimal solutions with drastically reduced computational time following extensive training on known network topologies~\cite{xu2023teal}.

Recently, deep learning methods have been explored to address the computational complexity challenges, a pioneering solution, {\it Teal}, incorporates graph neural network (GNN), multi-agent RL, and distributed optimization into the overall resilient TE solution workflow~\cite{xu2023teal}. It is shown to achieve near-optimal TE solutions with drastically reduced computational time following extensive training on known network topologies and traffic patterns.

%% file: future.tex
This study unveils critical aspects often overlooked in resilient WAN traffic engineering (TE). Our exploration highlights the importance of a holistic view of TE solutions to balance computational complexity, resource usage, and operational cost. Two new metrics are introduced—critical link set and network criticality, capturing demand imbalances and link utilization variations. 
They provide a more comprehensive assessment of resilient TE solutions. The proposed resilient TE enhancement, which makes the pre-computed tunnels adaptive to the 
traffic demand distribution and leverages proper solver selection, demonstrates promising efficiency gains across diverse WANs and traffic demand scales. More specifically, the adaptive tunnel scheme substantially cuts computation time while maintaining solution quality, making {\it congestion free} resilient TE solution feasible for large-scale networks.

While some performance metrics showed moderate degradation with the adaptive tunnel scheme in the big network, this tradeoff enables 
considerable improvements in computational efficiency and resource overprovisioning. Our future investigation will focus on more intelligent tunnel control strategies and more 
efficient optimization solver techniques that can better adapt to dynamic traffic demand distribution and WAN structure and capacity in real-world network environments. 
Expanding the solution to address a broader range of failure scenarios beyond single-link failures will also be a key task in future research.